\documentclass[%
 reprint,
 amsmath,amssymb,
 aps,
]{revtex4-2}

\usepackage{graphicx}
\usepackage{dcolumn}
\usepackage{bm}

\begin{document}

\preprint{APS/123-QED}

\title{Diffusion crossover from/to $q$-statistics to/from 
Boltzmann-Gibbs statistics  in the classical inertial $\alpha$-XY ferromagnet\\
}

\author{Antonio Rodr\'{\i}guez$^{1}$ and Constantino Tsallis$^{2,3,4}$}
\affiliation{%
 $^{1}$  GISC, Departamento de Matem\'atica Aplicada a la Ingenier\'{\i}a Aeroespacial, Universidad Polit\'ecnica de Madrid, Plaza Cardenal Cisneros 
s/n, 28040 Madrid, Spain\\
$^{2}$ Centro Brasileiro de Pesquisas F\'{\i}sicas and National Institute of Science and Technology for Complex Systems, Rua Dr. Xavier Sigaud 150, 22290-180, Rio de Janeiro, Brazil\\
$^{3}$ Santa Fe Institute, 1399 Hyde Park Road, Santa Fe, 87501 NM, United States\\
$^{4}$ Complexity Science Hub Vienna, Josefst\"adter Strasse 39, 1080 Vienna, Austria
}%

\date{\today}

\begin{abstract}
We study the angular diffusion 
in a classical $d-$dimensional inertial XY model with interactions decaying with the distance between spins as $r^{-\alpha}$, wiht $\alpha\geqslant 0$. After a very short-time ballistic regime, with $\sigma_\theta^2\sim t^2$, a super-diffusive regime, for which $\sigma_\theta^2\sim t^{\alpha_D}$, with $\alpha_D \simeq 1\text{.}45$ is observed, whose duration covers an initial quasistationary state and its transition to a second plateau characterized by the Boltzmann-Gibbs temperature $T_\text{BG}$.  Long after $T_\text{BG}$ is reached, a crossover to normal diffusion, $\sigma_\theta^2\sim t$, is observed. We relate, for the first time, via the expression $\alpha_D = 2/(3 - q)$, the anomalous diffusion exponent $\alpha_D$ with the entropic index $q$ characterizing the time-averaged angles and momenta probability distribution functions (pdfs), which are given by the so called $q-$Gaussian distributions, $f_q(x)\propto e_q(-\beta x^2)$, where $e_q (u) \equiv [1 + (1 - q)u]^{\frac{1}{1 - q}}$ ($e_1(u) = \exp(u)$).  For fixed size $N$ and large enough times, the index $q_\theta$ characterizing the angles pdf approaches unity, thus indicating a final relaxation to Boltzmann-Gibbs equilibrium. For fixed time and large enough $N$, the crossover occurs in the opposite sense.
\end{abstract}

\maketitle


\section{\label{introduction}Introduction} 
When a many-body system is said to be in {\it thermodynamical equilibrium}, two limits are, explicitly or implicitly, typically involved, namely the time $t \to\infty$ limit (stationary state) and the particle number $N\to\infty$ limit (large system). When we are dealing with a $d$-dimensional Hamiltonian system with say short-range two-body interactions (e.g., when coupling occurs locally, i.e., between neighboring elements, such as first neighbors), the energy probability distribution $p_l$ is given by the Boltzmann-Gibbs (BG) weight
\begin{equation}
p_l= \frac{e^{-\beta (E_l-\mu)}}{\sum_{l^{\prime}} e^{-\beta (E_{l^{\prime}}-\mu)}} \,,
\label{BG}
\end{equation}
where $\beta\equiv 1/kT$, $\mu$ is the chemical potential, and $\{E_l\}$ are the energies corresponding to the set $\{l\}$ of possible microscopic states. Under these circumstances, the ordering of the $t\to\infty$ and the $N\to\infty$ limits is irrelevant, i.e., we have an {\it uniform convergence} at the $1/t=1/N=0$ limiting point. It is generically so for all ensembles (micro-canonical, canonical, and grand-canonical). This remarkable uniform-convergence property and Eq. (\ref{BG}) are by no means warranted when we have long-range interactions (e.g., if the two-body coupling strength decays with distance $r$ like $1/r^\alpha$ with $0 \le \alpha <\infty$). Under such circumstances, Eq. (\ref{BG}) is generically false, the correct one appearing to be, in many cases, as follows \cite{Tsallis1988,Tsallis2023}:
\begin{equation}
p_l= \frac{e_q^{-\beta_q (E_l-\mu_q)}}{\sum_{l^{\prime}} e_q^{-\beta_q (E_{l^{\prime}}-\mu_q)}} \,,
\label{q}
\end{equation}
with
\begin{equation}
e_q^u\equiv [1+(1-q)u]^{\frac{1}{1-q}}_{+}  \;\;\;(q \in \mathbb{R};\,e_1^u=e^u) \,, 
\end{equation}
where $[\dots]_{+}=[\dots]$ if $[\dots] >0$ and zero otherwise. Clearly, Eq. (\ref{q}) recovers, for the particular instance $q=1$, Eq. (\ref{BG}).

The stationary-state probability distribution (\ref{q}) optimizes \cite{Tsallis1988,Tsallis2023}, under simple constraints, the entropic functional
\begin{eqnarray}
S_q &\equiv& k\frac{1-\sum_l p_l^q}{q-1} 
= 
k\sum_l p_l \ln_q \frac{1}{p_l } 
\nonumber \\
&=&
-k\sum_l p_l^q \ln_q p_l 
=
-k\sum_l p_l \ln_{2-q} p_l \;\;\;(q \in \mathbb{R})\,,
\end{eqnarray}
where $k$ is a positive conventional constant (in physics, $k = k_B$ and, in information theory, $k=1$), and $\ln_q z \equiv \frac{z^{1-q}-1}{1-q}\;(\ln_1 z = \ln z)$. It follows that
\begin{equation}
S_1 = S_{BG} \equiv -k \sum_l p_l \ln p_l \,.
\end{equation}
If the system $(A+B)$ is constituted by two subsystems $A$ and $B$ that are probabilistically independent, i.e., such that $p_{l,l^\prime}^{A+B}=p_l^A \,p_{l^\prime}^B, \forall (l,l^\prime)$, the definition of $S_q$ straightforwardly leads to
\begin{equation}
\frac{S_q(A+B)}{k}= \frac{S_q(A)}{k} + \frac{S_q(B)}{k} + (1-q)\frac{S_q(A)}{k}\frac{S_q(B)}{k} \,.  
\end{equation}
Therefore, unless $q=1$, $S_q$ is nonadditive (see \cite{Penrose1970} for the definition of entropic additivity). 

It is our present goal to illustrate that, when long-ranged two-body interactions are involved in a classical many-body Hamiltonian, crucial predictions of BG statistical mechanics are violated and we must replace them by their counterparts within the $q$-generalized theory (usually referred to as nonextensive statistical mechanics). In particular, the velocity distributions are, during longstanding intervals of time, not Maxwellians  but $q$-Gaussians instead. To be more precise, the regime corresponding to $N \gg 1$ (thermodynamical limit) and dimensionless time $t \gg 1$ (stationary-state limit) exhibits a crossover between BG behavior and $q$-statistical one. This corresponds to a {\it nonuniform convergence} at the $1/t=1/N=0$ limiting point. What essentially happens is that if $t$ approaches infinity quicker than $N$, the BG behavior prevails, whereas if $N$ approaches infinity quicker than $t$, the $q$-statistical behavior prevails. We illustrate these features through a {\it first-principle} molecular dynamical study of the $\alpha$-XY inertial ferromagnetic model.

\section{The $\alpha$-XY inertial model}

The $\alpha$-XY inertial model, introduced in \cite{AnteneodoTsallis1998},  consists of a set of 
$N$ interacting planar rotators governed by the Hamiltonian 
\begin{equation}
\mathcal{H} = K + V_\alpha = \frac{1}{2}\sum_{i = 1}^Np_i^2 + \frac{1}{2\tilde N}\sum_{j\neq i}\frac{1- \cos \left( \theta_i-\theta_j \right) }
{r_{ij}^\alpha}.\label{hamiltonian} 
\end{equation}
Each rotator, with unit moment of inertia, is characterized by a spin vector $\vec s_i = (\cos\theta_i,\sin\theta_i)$, so that interaction between rotators $i$ and $j$, given by $(1 - \vec s_i\cdot\vec s_j)/r_{ij}^\alpha$, decays with the distance between rotators $r_{ij} = |\boldsymbol{r}_i - \boldsymbol{r}_j|$ as a power law governed by the interaction range $\alpha\geqslant 0$. Rotators are arranged in positions $\boldsymbol{r}_i$ of a $d$-dimensional hypercubic lattice of linear size $L$, with $N  = L^d$, $d = 1, 2$ and 3, and unit lattice constant.

The scaling prefactor $\tilde N^{-1}$ in front of the potential term in Hamiltonian \eqref{hamiltonian}, with $\tilde N = \frac{1}{N}\sum_{i\neq j}r_{ij}^{-\alpha}$, is necessary in order to have a (conveniently) extensive total potential energy per particle \cite{JundKimTsallisPRB1995} for all values of $\alpha/d$. For $\alpha = 0$ (infinite-range regime), the thoroughly studied Hamiltonian Mean Field Model (HMF), introduced in \cite{AntoniRuffo1995}, is recovered. The opposite limiting case, i. e. $\alpha/d\to\infty$, corresponds to the nearest-neighbours (NN) approximation.  In between, there is the long-range regime ($1<\alpha /d < \infty$) as well as the very-long-range regime ($0<\alpha/d\leqslant 1$), case for which it can be shown \cite{Tamarit2000, Campa2000} that the same equation of state applies as for the HMF model, that is
\begin{equation}
U = {T \over 2} + \frac{1}{2} (1-M^{2})~, 
\label{equation_of_state}
\end{equation}
where $T$ is the temperature, $U$ the total energy per particle and the magnetization is given by $M = |\vec M|$, with $\vec M = \frac{1}{N}\sum_{i = 1}^N\vec s_i$. Equation of state \eqref{equation_of_state}, together with the consistency relation
\begin{equation}
M = \frac{I_1(M/T)}{I_0(M/T)},    
\label{consistency_relation}
\end{equation}
where $I_\mu(x)$ stands for the modified Bessel function, determines the existence of a paramagnetic-ferromagnetic continuous phase transition, which occurs at the critical point $(T_c, U_c) = (1/2,3/4)$, provided $0\leqslant\alpha/d\leqslant 1$.

Hamilton equations of motion for phase angles $\theta_i$ and conjugate momenta $p_i$ are easily derived from \eqref{hamiltonian} as

\begin{equation}
\dot{\theta}_i = \frac{\partial \mathcal{H}}{\partial p_i} = p_i~; \qquad 
\dot{p}_i = -\frac{\partial \mathcal{H}}{\partial \theta_i} 
=-\frac{1}{\widetilde{N}}\sum_{\substack{j\neq i}}^{\substack{N}}
\frac{\sin \left( \theta_i-\theta_j \right)}{r_{ij}^\alpha} \,. 
\label{hamilton_equations}
\end{equation}

Once solved, Eqs. \eqref{hamilton_equations} allow us to compute the kinetic temperature as $T(t) = \frac{2K(t)}{N}$, as well as the mean squared displacement
\begin{equation}
\sigma^2(t) \equiv \frac{1}{N}\sum_{i = 1}^N(\theta_i(t) - \theta_i(0))^2.
\label{mean_squared_displacement}
\end{equation}
The total energy per particle $U = \frac{1}{N}[K(t) + V_\alpha(t)]$ as well as the total momentum $P = \sum_{i = 1}^Np_i(t)$ are constants of the motion.

\section{Anomalous to/from normal diffusion crossover as a function of $t$ and $N$ .}

We have integrated the equations of motion \eqref{hamilton_equations} by means of a fourth-order symplectic algorithm \cite{Yoshida} and an integration step $h = 0\text{.}2$, which is small enough to keep the total energy per particle constant throughout all the calculations within a relative error $\varepsilon = 10^{-5}$. Periodic boundary conditions have been considered so that a Fast Fourier Transform algorithm has been used to efficiently compute the interactions. In order to compare equal number of rotators for different dimensions, we have considered system sizes as sixth powers: $N = (a^6)^1 = (a^3)^2 = (a^2)^3$, with $a = 4, 5, 6$ and 7. As initial conditions we have taken angles from a uniform distribution in $[-\pi, \pi]$ so that the initial total magnetization is $\vec M(0) \sim \vec 0$. The initial momenta have been taken from a symmetric uniform distribution, then shifted to get total momentum $P = 0$, and rescaled to get the desired total energy per particle $U$.   

\begin{figure}[h]
\includegraphics[width = 10 cm, clip =]{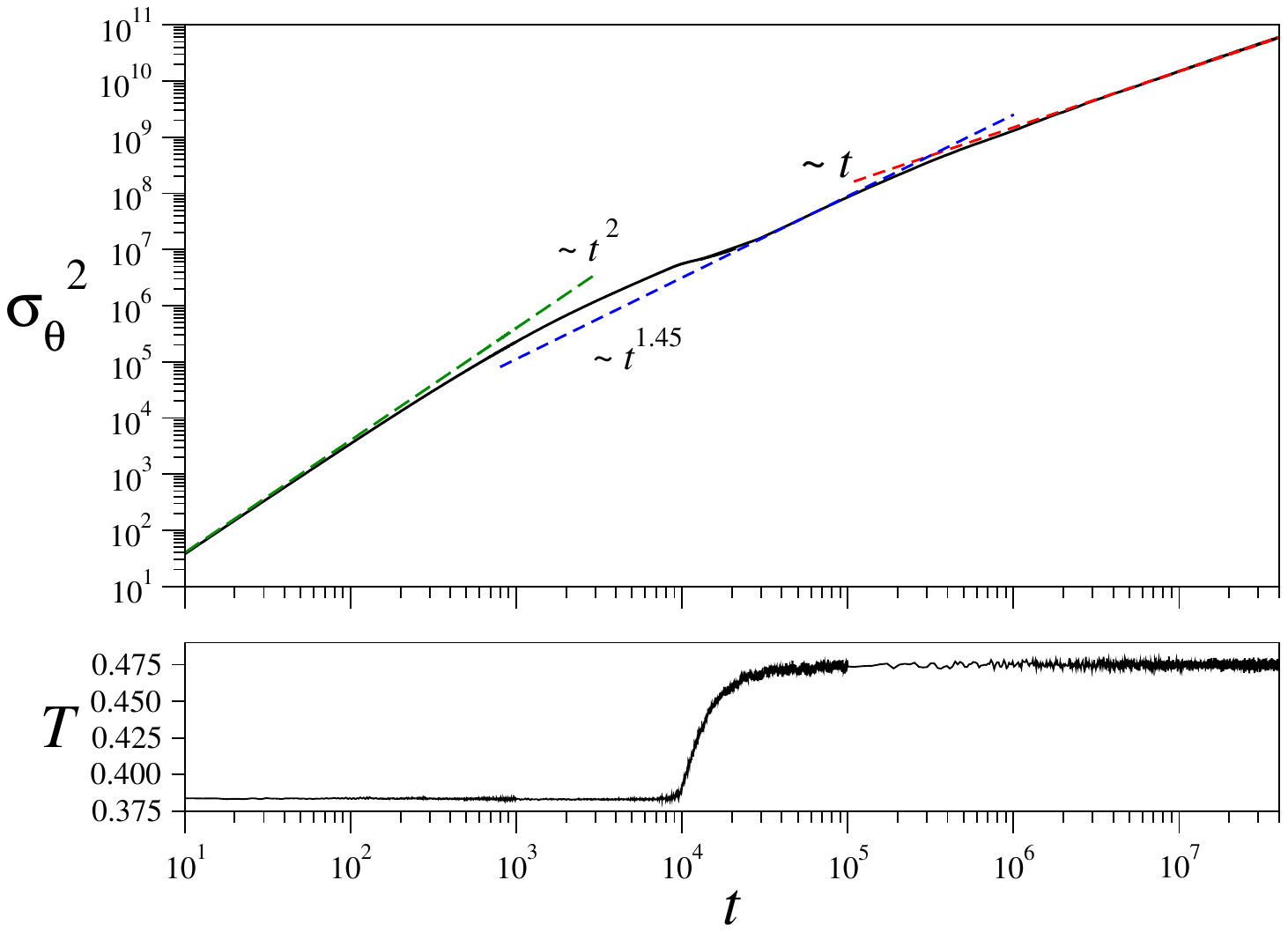}
\caption{Mean squared displacement \eqref{mean_squared_displacement} (top panel) and temperature (bottom panel) versus time for a system with $N = 7^6$, $\alpha = 0.9$ and $d = 1$. For short times, up to $t\sim 10^2$, a ballistic regime is observed. Then, the system enters the superdiffusive regime coinciding with the temperature transition from $T_\text{QSS}$ to $T_\text{BG}$. Later on, the system reaches normal diffusion after the characteristic time $t_\text{c} \simeq  309000$, calculated as the intersection point of asymptotic lines $\sigma^2\sim t^{1.45}$ and $\sigma^2\sim t$.}
\label{fig1}
\end{figure}

Fig. \ref{fig1} shows the time evolution of both the temperature and the mean squared displacement \eqref{mean_squared_displacement} of a one dimensional system of $N = 7^6 = 117649=343^2=49^3$ rotators with interaction range $\alpha = 0\text{.}9$ and total energy per particle $U = 0\text{.}69$. The temperature curve shows the transition from the initial quasi-stationary state (QSS), with a temperature $T_\text{QSS}\simeq 2U - 1$ (given by the extension to lower energies of the linear part of the caloric curve \cite{RodriguezNobreTsallis_PRE_105}), to the equilibrium Boltzmann-Gibbs (BG) temperature, $T_\text{BG}$,  given by the equation of state \eqref{equation_of_state}. At very short times, the mean squared displacement scales with time as $\sigma_\theta^2\sim t^2$, so the usual ballistic regime is obtained~\cite{LatoraRapisardaRuffo}. When the transition to the $T_\text{BG}$ state is approached, the diffusion changes to an anomalous regime, $\sigma_\theta^2\sim t^{\alpha_D}$, with $\alpha_D \simeq 1\text{.}45$. Only long after $T_\text{BG}$ has been reached, the system enters the normal diffusion regime $\sigma_\theta^2\sim t$. Thus, besides the characteristic time $t_\text{QSS} \simeq 12930$ for which the system transitions from the QSS to the $T_\text{BG}$ state, a second characteristic time, $t_\text{c} \simeq 309000$ (see details in caption of Fig. \ref{fig1}) appears, for which the transition from supernormal to normal diffusion takes place.   

\begin{figure}[h]
\includegraphics[width = 13 cm, clip =]{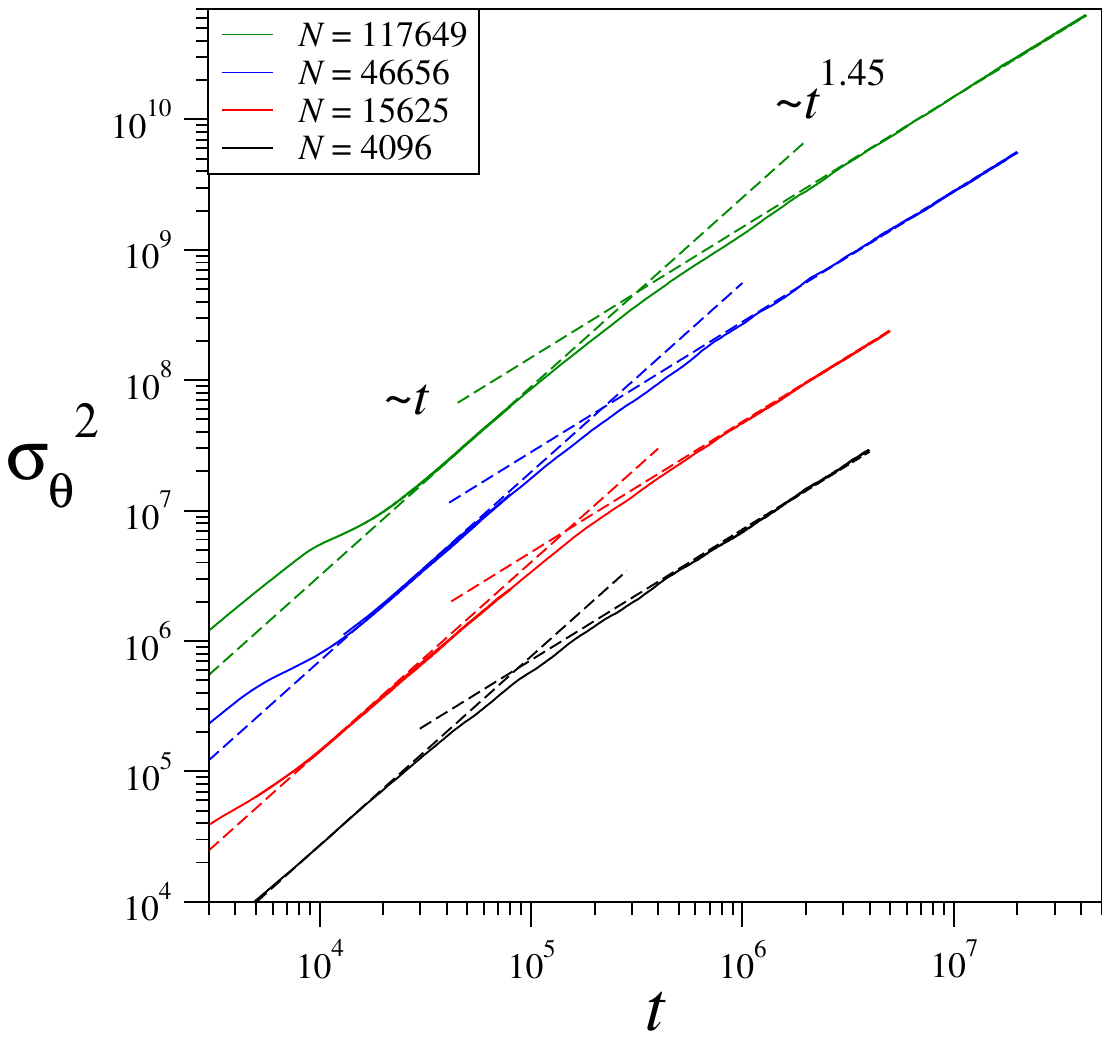}
\caption{Mean squared displacement versus time curves for systems with $\alpha = 0.9$, $d = 1$, $U = 0\text{.}69$ and $N = 4^6, 5^6, 6^6$ and $7^6$. For better visibility, we have rescaled by successive factors 1/4 the ordinates of $N = 6^6$, $5^6$ and $4^6$ curves (the $N = 7^6$ curve has no shift).}
\label{fig2}
\end{figure}

A study of the duration $t_\text{QSS}$ of the QSS as a function of $N$ and $\alpha/d$ was made in \cite{RodriguezNobreTsallis2021a} where it was shown that it follows the scaling law $t_\text{QSS}\sim N^{A(\alpha/d)}$. Thus, as $t_\text{QSS}$ grows with the system size, it is also expected that $t_\text{c}$ follows the same trend. In Fig. \ref{fig2} we show again Fig. \ref{fig1} results for the time evolution of $\sigma_\theta^2$, together with those characteristic times corresponding to system sizes $N = 4^6 = 4096$, $5^6 = 15625$ and $6^6 = 46656$, with the same parameter values used in Fig. \ref{fig1}. A clear increase of $t_\text{c}$ with $N$ is observed.

\begin{figure}[h]
\includegraphics[width = 10 cm, clip =]{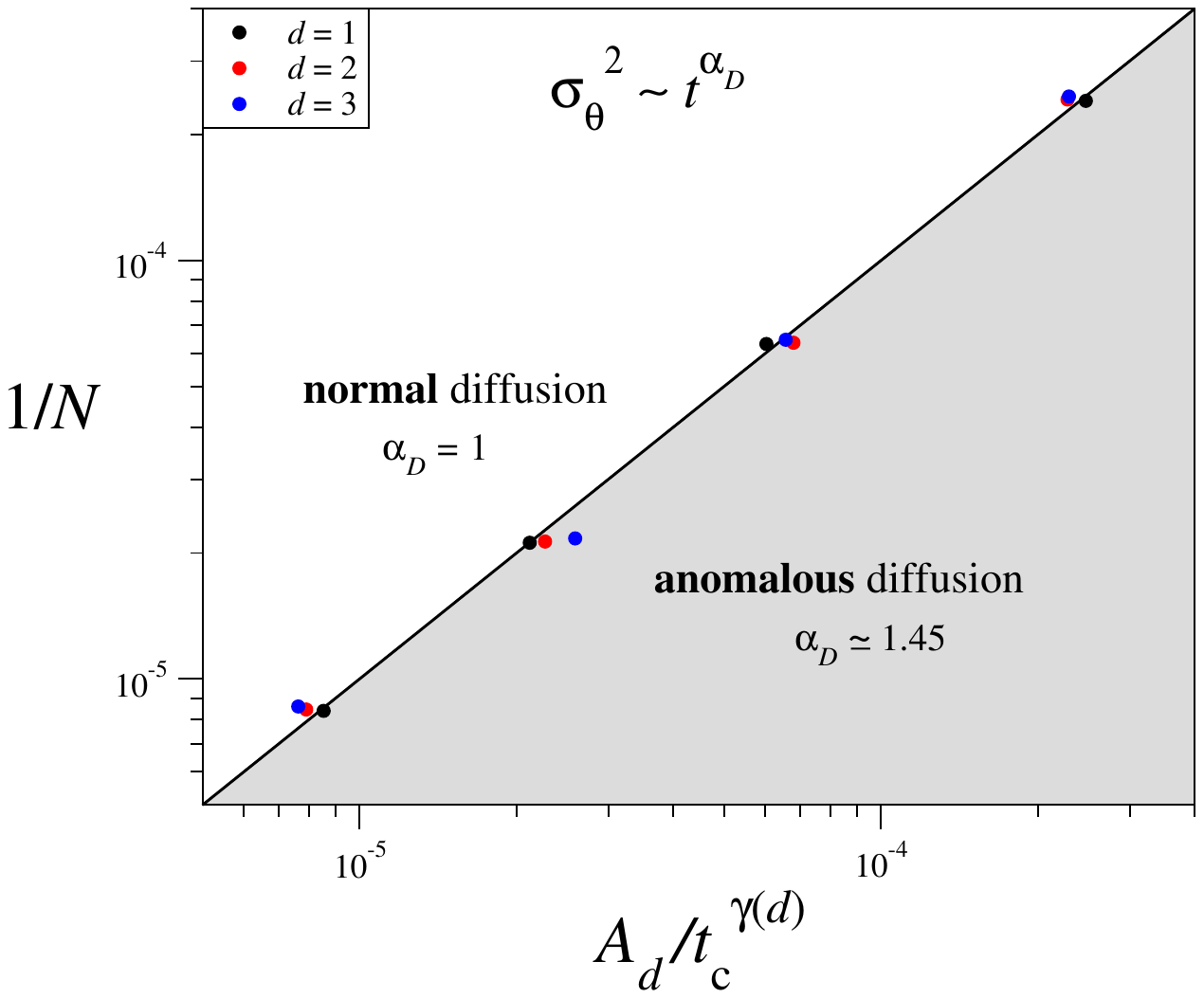}
\includegraphics[width = 10.5 cm, clip =]{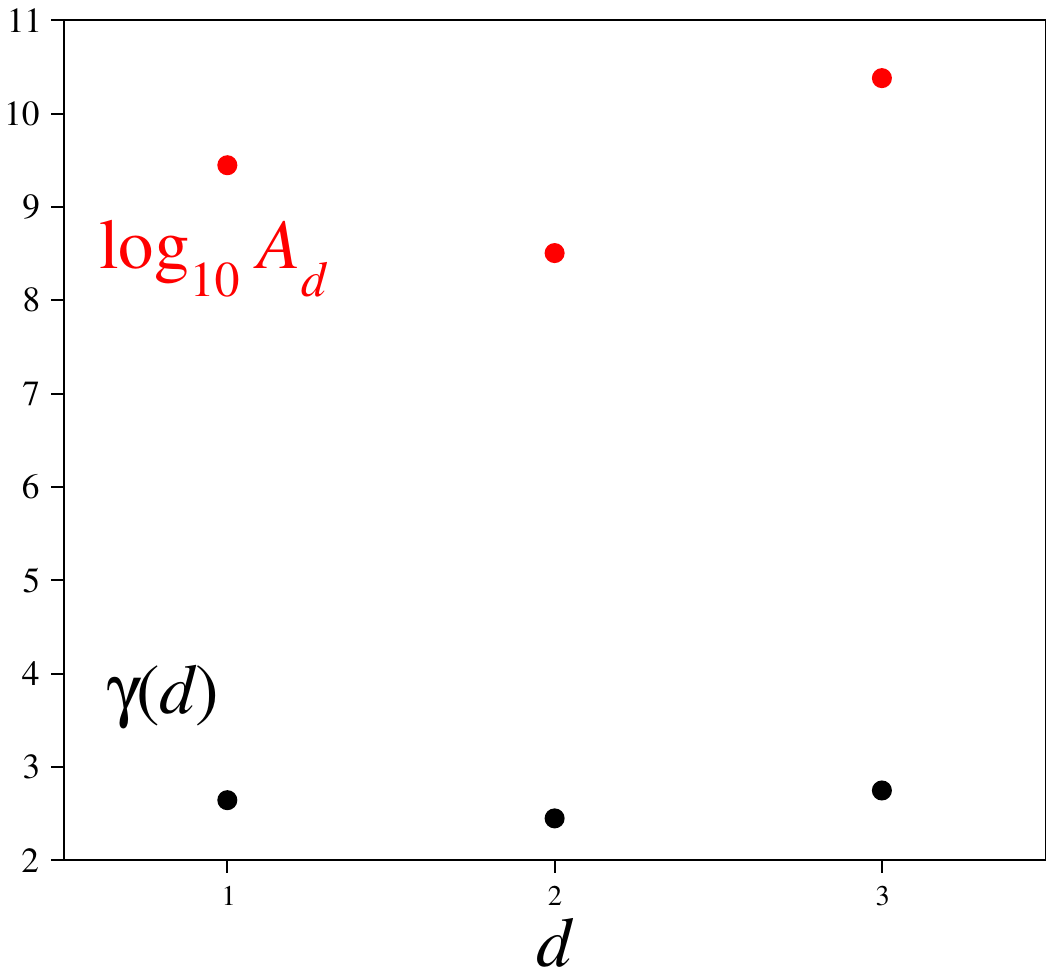}
\caption{{\it Top}: Phase diagram showing normal and anomalous diffusion regimes. The bisecting line separating both regimes corresponds to the fitting of the $t_\text{c}(N, d)$ curves as $A_dN = t_\text{c}^{\gamma(d)}$, for $\alpha/d = 0\text{.}9$,  $d = 1, 2$ and 3 and $N = 4^6, 5^6, 6^6$ and $7^6$. {\it Bottom}: The values of the overlap parameters are $(A_1, \gamma(1)) = (2\text{.}8\times 10^9, 2\text{.}644)$, ($A_2, \gamma(2)) = (3\text{.}2\times 10^8, 2\text{.}449$) and ($A_3, \gamma(3)) = (2\text{.}4\times 10^{10}, 2\text{.}748$)}
\label{fig3}
\end{figure}

Results that are not shown here indicate an increase of $t_\text{c}$ with the spatial dimension $d$ (as it is the case for $t_\text{QSS}$ \cite{RodriguezNobreTsallis2021a}). Our simulations confirm a scaling relation of the characteristic time $t_\text{c}(N, d)$ with the system size and the spatial dimension in the form $N = A_d^{-1}t_\text{c}^{\gamma(d)}$. In Fig. \ref{fig3} we show, in reciprocal ($N^{-1}, A_dt_\text{c}^{-\gamma(d)})$ coordinates, $t_c$ values extracted from Fig. \ref{fig2} together with those corresponding to dimensions $d = 2$ and 3, with $\alpha/d = 0\text{.}9$ and equal values of $N$. We have collapsed the three $t_\text{c}(N, d)$ curves, corresponding to dimensions $d = 1$, 2 and 3, into a single one, thus drawing a line separating two distinct regimes, corresponding to normal and anomalous diffusion in a phase diagram. As a result, the thermodynamic ($N\to\infty$) and stationary-state $t\to\infty$ limits are not exchangeable. Fixing the value of $N$ and waiting long enough the system ends up in the normal diffusion regime (upper part of the phase diagram), while taking first the thermodynamic limit means staying for ever in the anomalous diffusion regime (lower part of the phase diagram).

We will show next that superdiffusion is closely related to other anomalous feature already studied in the $\alpha-$XY model. It has been shown  \cite{CirtoAssisTsallis2014, CirtoRodriguezNobreTsallis2018} that time averaged angles as well as momenta probability distributions of the $\alpha-$XY model correspond with $q-$Gaussians (meaning $y \propto e_q^{-b \, x^2}$), a generalization of the Gaussian distributions, which extremize, under simple constraints, the $q$-entropy $S_q$, the entropic functional which is at the basis of nonextensive statistical mechanics \cite{Tsallis1988,Tsallis2023}. The relation 
\begin{equation}
\alpha_D = \frac{2}{3 - q}    
\label{q-alpha}
\end{equation} 
between the entropic index $q$ and the diffusion exponent $\alpha_D$ was established in \cite{TsallisBukman1996}, where driven anomalous diffusion described by a nonlinear Fokker-Plank-like equation, namely the Plastino-Plastino equation \cite{PlastinoPlastino1995}, was studied. The relation (\ref{q-alpha}) has been repeatedly verified experimentally in diverse complex systems, and notably for granular matter \cite{CombeRichefeuStasiakAtman2015}.  

\begin{figure}[h]
\includegraphics[width = 12 cm, clip =]{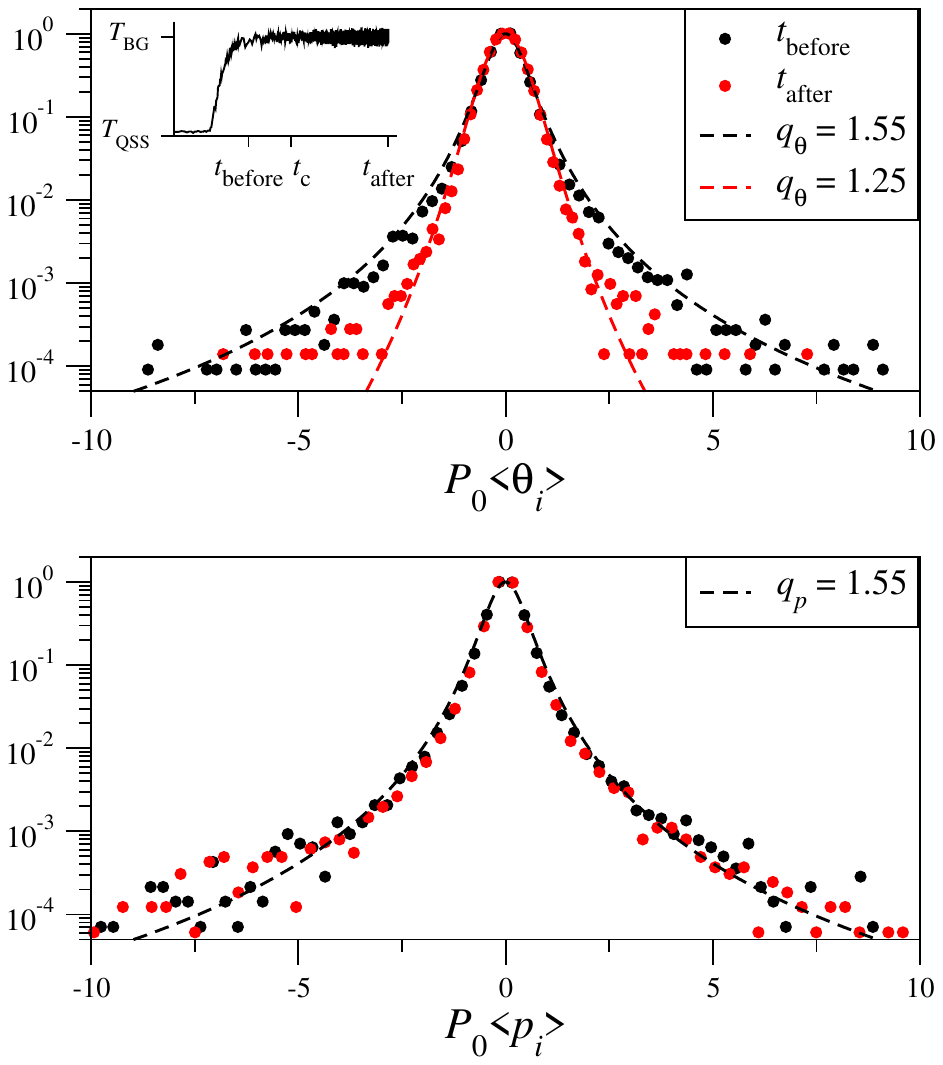}
\caption{Top panel: pdfs for time averaged angles for a system with $N = 6^6$, $\alpha = 0.9$ and  $d = 1$ within the time intervals $[t_\text{before}, t_\text{before} + \Delta t]$ (black bullets) and $[t_\text{after}, t_\text{after} + \Delta t]$ (red bullets) with $\Delta t = 5\times10^3$, which correspond to $q_{\theta}$-Gaussians with $q_\theta \simeq 1\text{.}55$ and 1.25 respectively. The inset shows, in linear-log coordinates, the time evolution of the temperature, showing $t_\text{before} = 31\times 10^3$, $t_\text{c} \simeq  219\times 10^3$ and $t_\text{after} = 19931\times 10^3$. Bottom panel: pdfs for the time averaged momenta and the same parameter values as the top panel. The entropic index of the $q$-Gaussian for both time windows is the same: $q_p \simeq 1\text{.}55$.}  
\label{fig4}
\end{figure}

Fig. \ref{fig4} shows time averaged angles (top panel) as well as momenta (bottom panel) probability distribution functions for a one-dimensional system of $6^6$ rotors with interaction range $\alpha = 0\text{.}9$ and total energy per particle $U = 0\text{.}69$. In both cases a time window of width $50\times 10^3$ time units has been used for averaging. When the time window starts at $t_\text{before} = 31\times 10^3$ --- that is, already in the $T_{BG}$ regime though before the characteristic time $t_\text{c} \simeq 219\times 10^3$, hence still in the anomalous diffusion regime --- similar $q$-Gaussians are obtained for both the time averaged angles and momenta pdfs, with equal entropic index $q_\theta \simeq q_p \simeq  1\text{.}55$, which, when introduced in \eqref{q-alpha} yields $\alpha_D \simeq  1\text{.}38$, a value which, within a 5\% relative error, is consistent with  $\alpha_D \simeq  1\text{.}45$ obtained in our simulations. This value of the entropic index persists for a very long time after the transition to the normal diffusion regime has occurred.  Nevertheless, when waiting long enough, $q_\theta$ starts to decrease approaching $q_\theta = 1$, as can be seen in the top panel of Fig. \ref{fig4} when the time window is opened at $t_\text{after} = 19931\times 10^3$. Nevertheless, $q_p$ remains unaltered even at such very large time. For fixed $N$, a complete relaxation to Boltzmann equilibrium, that is both $q_\theta\to 1$ and $q_p\to 1$, is expected for larger times, still unattainable for our current computing capacity.  


\section{Conclusions}

We have numerically studied the first-principle dynamics of a long-range-interacting many-body classical Hamiltonian system, namely the $d$-dimensional $\alpha$-XY inertial ferromagnetic model. 
Our focus has been the angular crossover which occurs from/to $q$-statistical (anomalous superdiffusion) to/from BG (normal diffusion) behaviors. For fixed $N$ and increasing time $t$, the crossover occurs from $q$-statistics to BG statistics. For fixed time and increasing $N$, the crossover occurs in the opposite sense, i.e, from BG statistics to $q$-statistics. Therefore, nonuniform convergence is observed at the $1/N=1/t=0$ limiting point, as illustrated in Fig. \ref{fig3}. These results concretely exhibit, for a $d=1,2,3$ paradigmatic long-range-interacting model with $\alpha/d=0.9$, the nonuniform convergence that was in fact conjectured long ago (see \cite{Tsallis1999}). For fixed $N$, the crossover from $q$-statistics to BG statistics does not occur at the same time for different relevant quantities: it first occurs for the time-dependence of the angular variance (from anomalous to normal diffusion), later on for the entire angle distribution of probabilities, and only finally (presumably), at even larger times, for the momenta distribution of probabilities. Let us also mention that the scaling relation (\ref{q-alpha}) is numerically satisfied. 

The present results are consistent with the preliminary ones observed in the long-range-interacting Fermi-Pasta-Ulam-Tsingou-like model \cite{ChristodoulidiTsallisBountis2014}. Further analytical and/or numerical studies of the present crossover -- apparently quite general -- are very welcome.

\begin{acknowledgments}
We have benefited from partial financial support by CNPq and Faperj (Brazilian agencies).
\end{acknowledgments}


\begin{thebibliography}{99}

\bibitem{Tsallis1988}C. Tsallis, Possible generalization of Boltzmann-Gibbs statistics,  J. Stat. Phys. {\bf 52}, 479-487 (1988).

\bibitem{Tsallis2023} 
C. Tsallis, {\em Introduction to Nonextensive Statistical Mechanics: Approaching a Complex
World}, (Springer, 2009); Second Edition (Springer, 2023).

\bibitem{Penrose1970}O. Penrose, {\it Foundations of Statistical Mechanics: A Deductive Treatment} (Pergamon, Oxford, 1970), page 167.

\bibitem{AnteneodoTsallis1998}C. Anteneodo and C. Tsallis, Breakdown of the exponential sensitivity to the initial conditions: Role of the range of the interaction, Phys. Rev. Lett. {\bf 80}, 5313 (1998).


\bibitem{JundKimTsallisPRB1995}
P. Jund, S.G. Kim and C. Tsallis, 
Crossover from extensive to nonextensive behavior driven by long-range interactions,
Phys. Rev. B {\bf 52}, 50 (1995). 

\bibitem{AntoniRuffo1995}M. Antoni and S. Ruffo,  Clustering and relaxation in Hamiltonian long-range dynamics, Phys. Rev. E {\bf 52}, 2361-2374 (1995).

\bibitem{Tamarit2000}
F. Tamarit and C. Anteneodo, 
Rotators with long-range interactions: Connection with the mean-field 
approximation, 
Phys. Rev. Lett. {\bf 84}, 208 (2000). 


\bibitem{Campa2000}
A. Campa, A. Giansanti and D. Moroni, 
Canonical solution of a system of long-range interacting rotators on a lattice, 
Phys. Rev. E {\bf 62}, 303 (2000). 


\bibitem{Yoshida} H. Yoshida. Construction of higher order symplectic integrators, Phys. Let. A {\bf 150}, 262 (1990).


\bibitem{RodriguezNobreTsallis_PRE_105} A. Rodr\'{\i}guez, F.D. Nobre and C. Tsallis,  Finite-size scaling of quasi-stationary-state temperature,  Phys. Rev. E {\bf 105}, 044111 (2022).


\bibitem{LatoraRapisardaRuffo} V. Latora, A. Rapisarda and S. Ruffo, Superdiffusion and out-of-equilibrium chaotic dynamics with many degrees of freedom. Phys. Rev. Lett. {\bf 83}, 2104 (1999).

\bibitem{RodriguezNobreTsallis2021a}A. Rodr\'{\i}guez, F.D. Nobre and C. Tsallis, Quasi-stationary-state duration in the classical $d$-dimensional long-range inertial XY ferromagnet, Phys. Rev. E {\bf 103}, 042110 (2021).


\bibitem{CirtoAssisTsallis2014}L.J.L. Cirto, V.R.V. Assis and C. Tsallis, Influence of the interaction range on the thermostatistics of a classic many-body system, Physica A {\bf 393}, 286 (2014). 

\bibitem{CirtoRodriguezNobreTsallis2018}L.J.L. Cirto, A. Rodr\'{\i}guez, F.D. Nobre and C. Tsallis, Validity and failure of the Boltzmann weight, EPL {\bf 123}, 30003 (2018). 


\bibitem{TsallisBukman1996} C. Tsallis and D. J. Bukman,  Anomalous diffusion in the presence of external forces -  Exact time-dependent solutions and their thermostatistical basis, Phys. Rev. E {\bf 54}, R2197 (1996).

\bibitem{PlastinoPlastino1995}A.R. Plastino and A. Plastino, Non-extensive statistical mechanics and generalized Fokker-Planck equation, Physica A  {\bf 222}, 347 (1995).

\bibitem{CombeRichefeuStasiakAtman2015}G. Combe, V. Richefeu, M. Stasiak and A.P.F. Atman, Experimental validation of nonextensive scaling law in confined granular media, Phys. Rev. Lett. {\bf 115}, 238301 (2015).

\bibitem{Tsallis1999}C. Tsallis, Nonextensive statistics: Theoretical, experimental and computational evidences and connections, in {\it Nonextensive Statistical Mechanics and Thermodynamics}, eds. S.R.A. Salinas and C. Tsallis, Braz. J. Phys. {\bf 29} (1), 1-35  (1999). 

\bibitem{ChristodoulidiTsallisBountis2014}H. Christodoulidi, C. Tsallis and T. Bountis, Fermi-Pasta-Ulam model with long-range interactions: Dynamics and thermostatistics, EPL {\bf 108}, 40006 (2014).


\end{thebibliography}
\end{document}